\documentclass[article,twocolumn, floatfixnofootinbib]{revtex4}
\usepackage{filecontents}
\usepackage{hyperref}
\usepackage{graphicx} 
\usepackage{subfig}
\usepackage{color}

\newcommand{\ket}[1]{\left| #1 \right\rangle}
\newcommand{\bra}[1]{\left\langle #1 \right|}

\newcommand{\be}{\begin{equation}}
\newcommand{\ee}{\end{equation}}
\newcommand{\bea}{\begin{eqnarray}}
\newcommand{\eea}{\end{eqnarray}}

\usepackage{dcolumn}
\usepackage{bm}
\usepackage{amssymb,amsmath}
\usepackage{color}
\usepackage{float}
\usepackage{tikz}
\usetikzlibrary{arrows}

\definecolor{DarkGreen}{rgb}{0,0.6,0.2}

\begin{document}
\title{Fano-Agarwal couplings and non-rotating wave approximation in single-photon timed Dicke subradiance}
\author{$^{1,2}$Imran M. Mirza and $^{1}$Tuguldur Begzjav }
\affiliation{ $^{1}$Department of Physics, Texas A\&M University, College Station, TX 77843, USA\\
$^{2}$Department of Physics, University of Michigan, Ann Arbor, MI 48109-1040, USA}
\begin{abstract}
Recently a new class of single-photon timed-Dicke (TD) subradiant states has been introduced with possible applications in single-photon based quantum information storage and on demand ultrafast retrieval \cite{scully2015single}. However, the influence of any kind of virtual processes on the decay of these new kind of subradiant states has been left as an open question. In the present paper, we focus on this problem in detail. In particular, we investigate how pure Fano-Agarwal couplings and other virtual processes arising from non-rotating wave approximation impact the decay of otherwise sub- and superradiant states. In addition to the overall virtual couplings among all TD states, we also focus on the dominant role played by the couplings between specific TD states. 
\end{abstract}
\maketitle
\section{Introduction}

Cooperative spontaneous emission from an atomic ensemble has been a well-established problem of interest in quantum optics since the pioneering work of Dicke in 1954 \cite{dicke1954coherence}. The subject has gained renewed interest in recent years when the problem of single-photon absorption by a collection of resonant two-level atoms is considered \cite{scully2006directed,svidzinsky2008dynamical,scully2009super,sete2010correlated,svidzinsky2015quantum}. 
The consequential phenomenon of single-photon superradiance is a pure quantum many-body effect in which system evolves into an entangled state \cite{wiegner2011quantum} and real and virtual photons are exchanged among different atoms in the ensemble through the interaction field. The inclusion of virtual transitions and Lamb shifts in the problem turns out to bring fascinating insights related to the field of quantum electrodynamics \cite{scully2009collective,scully2010lamb,friedberg2008effects}. Accordingly, in the last decade single-photon superradiance has witnessed a flurry of research activity with a wide range of applications in quantum optics, quantum information and condensed matter physics\cite{scully2006directed,eisaman2004shaping,kalachev2007quantum,svidzinsky2013quantum,kuzmich2003generation,
wang2015superradiance,porras2008collective,wang2015topological}.\\
In the standard treatment of the problem a collection of N identical two-level atoms with a ground (excited) state $\ket{b}$ ($\ket{a}$) with transition frequency $E_{a}-E_{b}=\hbar\omega$ is considered. When a single photon is collectively absorbed, the system forms a superposition state in which $\beta_{i}(r_{ij},t)$ describes the probability amplitude associated with the $i$th member of the state ($r_{ij}=|r_{i}-r_{j}|$ is the inter-atomic separation between the $i$th atom and some reference atom $j$). If the virtual processes, polarization of light and retardation effects are ignored, the resultant time evolution of $\beta_{i}(t)$ under the Markov approximation follows \cite{svidzinsky2012nonlocal,ernst1969coherent,feng2014effect}:
\begin{equation}
\frac{\partial\beta_{i}(t,r_{ij})}{\partial t}=-\frac{\gamma}{N}\sum^{N}_{j=1}\frac{sin(k_{0}|r_{i}-r_{j}|)}{(k_{0}|r_{i}-r_{j}|)}\beta_{j}(t,r_{ij}),
\end{equation}
here $\gamma$ is the single atom decay rate and $k_{0}=\omega/c$ with $c$ being the speed of light. 
The problem becomes even more richer when the virtual processes are also incorporated. In that case it is known that the time evolution of $\beta_{i}(t,r_{ij})$ involves an exponential kernel rather than a sine kernel \cite{friedberg1973frequency}
\begin{equation}
\frac{\partial\beta_{i}(t,r_{ij})}{\partial t}=\frac{i\gamma}{N}\sum^{N}_{j=1}\frac{exp(ik_{0}|r_{i}-r_{j}|)}{(k_{0}|r_{i}-r_{j}|)}\beta_{j}(t,r_{ij}).
\end{equation}
The single-photon ``timed" Dicke (TD) state ($\ket{+}_{\vec{k}_{0}}$) was first introduced by Scully et.al in references \cite{scully2006directed,scully2007correlated}:
\begin{equation}
\ket{+}_{\vec{k}_{0}}=\frac{1}{\sqrt{N}}\sum^{N}_{j=1}e^{i\vec{k}_{0}.\vec{r}_{j}}\ket{b_{1}b_{2}...a_{j}...b_{N}},
\end{equation}
where $\ket{b_{1}b_{2}...a_{j}...b_{N}}$ is a Fock state in which atom at the $j$th location is excited and all other atoms are in the ground state. On contrary to the ordinary Dicke (OD) state :$\ket{+}=\frac{1}{\sqrt{N}}\sum^{N}_{j=1}\ket{b_{1}b_{2}...a_{j}...b_{N}},$ in the TD states the atoms in the ensemble are excited at different times depending on their positions in the ensemble (as manifested by the phase factors which can also be interpreted as the timing factors through $t_{j}=\vec{k}_{0}.\vec{r}_{j}/\omega$).\\

Despite of extensive work on the single-photon superradiance, single-photon subradiance has been less studied essentially due to weak interaction of subradiant states with the environment and their elevated sensitivity on non-radiative damping processes. However, recently subradiance has started to gather both theoretical and experimental attention \cite{scully2015single,guerin2016subradiance,mcguyer2015precise} due to their promising applications in quantum information storage. In this context, M. O. Scully has introduced and analyzed a new class of TD subradiant states \cite{scully2015single}. The first member in this class $\ket{-}_{\vec{k}_{0}}$ state is expressed as:
\begin{equation}
\begin{split}
&\ket{-}_{\vec{k}_{0}}=\\
&\frac{1}{\sqrt{2N_{2}}}\sum^{2}_{j,j^{'}}\Bigg(e^{i\vec{k}_{0}.\vec{r}_{j}}\ket{a_{j}b_{j^{'}}}-e^{i\vec{k}_{0}.\vec{r}_{j^{'}}}\ket{b_{j}a_{j^{'}}}\Bigg)\ket{\lbrace b_{j}b_{j^{'}}\rbrace},
\end{split}
\end{equation}
primed and unprimed indices mark the atoms belonging to two different sections of the sample. As emphasized in Ref.\cite{scully2015single} the basic motivation of generating $\ket{-}_{\vec{k}_{0}}$ state is to utilize it for single-photon storage on the time scale shorter than $\gamma^{-1}$. However, any kind of virtual processes and their influence on the decay of this (and other) new kind of subradiant states has not been left as an open question in that paper. In the present work, we focus on this problem in detail.\\

There are two types of virtual processes we'll address in this context: (1) Fano-Agarwal (FA) couplings that arise due to the interaction of discrete atomic energy levels with environmental mode continuum and (2) Virtual processes arising from the non-rotating wave approximation (NRWA).
In the presence of pure FA couplings, we find that the individual populations tends to achieve smaller highest values as the number of atoms in the atomic ensemble are increased, however the summed up effect of all FA couplings remains substantial. Moreover, we notice that the dominant FA coupling between individual TD states depend on the initial state of the system. Finally, the inclusion of scalar Lamb shift and NRW terms yields a small effect on the decay of $\ket{+}_{\vec{k}_{0}}$ state, while $\ket{-}_{\vec{k}_{0}}$ (and other TD subradiant states) show markedly fast decay.\\

In the next section, we'll start off by introducing the system model and a transformation between the Fock state basis and the TD basis is presented. 
\section{System Hamiltonian and the transformation between timed Dicke and Fock state basis}
Following the paradigm model, we consider an atomic ensemble of identical two level atoms coupled to a single environment. The environment/bath is modeled to have a continuum of modes where frequency of the $k$th mode is represented by $\nu_{k}$. The interaction picture Hamiltonian of the system is expressed as:
\begin{equation}
\begin{split}
&\hat{V}=\sum^{N}_{j=1,k}g_{k} [(\hat{\sigma}_{j}e^{-i\omega t}+\hat{\sigma}^{\dagger}_{j}e^{i\omega t})( \hat{a}^{\dagger}_{\vec{k}}e^{i\nu_{k}t-i\vec{k}.\vec{r}_{j}}\\
& +\hat{a}_{\vec{k}}e^{-i\nu_{k}t+i\vec{k}.\vec{r}_{j}})],
\end{split}
\end{equation}
where $\hbar=1$ and $\vec{r}_{j}$ is the position vector of the $j$th atom in the ensemble. $g_{k}=(\mathcal{P}/\hbar)\sqrt{\hbar\nu_{k}/\epsilon_{0}V}$ is the atom-environment coupling rate with $\mathcal{P}$ being the dipole moment matrix element, $V$ is the volume of the sample and $\epsilon_{0}$ is the permittivity constant. The reason for not making the RWA in Eq.~5 is a known fact that RWA leads to an improper treatment of the virtual processes \cite{svidzinsky2010evolution,li2012collective}. The annihilation of the photon in the $k$th environmental mode is described by the operator $\hat{a}_{k}$ and $\hat{\sigma_{j}}$ is the lowering operator for the jth atom. Non-vanishing commutation relations are:
\begin{equation*}
[\hat{a}_{k},\hat{a}^{\dagger}_{k^{'}}]=\delta_{kk^{'}}, [\hat{\sigma}_{j},\hat{\sigma}^{\dagger}_{l}]=\hat{\sigma}_{z}\delta_{jl}, \forall \lbrace j,l \rbrace=1,2,...,N.
 \end{equation*}
 
The state describing the single excitation in the global system (atoms plus the field) in the timed-Dicke (TD) basis can be expressed as:
\begin{equation}
\begin{split}
&\ket{\Psi(t)}=\Bigg(\beta_{+}\ket{+}_{\vec{k}_{0}} + \beta_{-}\ket{-}_{\vec{k}_{0}} +...+\beta_{-}\ket{N}_{\vec{k}_{0}}\Bigg)\otimes \ket{0}\\
&+\sum_{\vec{k}}\gamma_{\vec{k}}(t)\ket{b_{1},b_{2},...,b_{N},1_{\vec{k}}}\\
&+\sum_{\vec{k},i,j}\eta_{\vec{k}}(t)e^{ik_{0}.\vec{r}_{ij}}\ket{b_{1},b_{2},...,a_{i},a_{j},...,b_{N},1_{\vec{k}}}
\end{split}
\end{equation}
Notice that this states now includes second order/two photon processes as well as exhibited by the term with amplitude $\eta_{\vec{k}}(t)$. This term describes a situation in which both atoms in the ensemble are excited and there is one (virtual) photon in the field with ``negative" energy.
$\ket{0}$ and $\ket{1_{\vec{k}}}$ are the environment states with zero and one photon in the $k$th mode of the continuum, respectively.
 It turns out for ensemble with larger number of atoms, the choice of TD states as a basis makes the problem intricate for both analytic and numerical solutions. In view of this, we introduce a basis transformation. Suppose we solve the present problem in a Fock (bare) basis first. The final expression of time-evolution can be represented as:
\begin{equation}
\frac{\partial\mathcal{B}_{b}}{\partial t}=\mathcal{M}\mathcal{B}_{b}
\end{equation}
here $\mathcal{B}_{b} = (\beta_{1} \beta_{2} ... \beta_{N})^{T}$ is the bare/Fock basis column matrix. $\mathcal{M}$ is a square matrix which depends on the system parameters (for instance inter-atomic separations). Next we introduce the transformation through the unitary matrix $\mathcal{S}$ as: $\mathcal{B}_{TD}=\mathcal{S}\mathcal{B}_{b}$ such that:
\begin{equation}
\frac{\partial\mathcal{B}_{TD}}{\partial t}=\mathcal{S}\mathcal{M}\mathcal{S}^{-1}\mathcal{B}_{TD}
\end{equation}
\section{\bf {Pure Fano-Agarwal couplings}}  
When the problem of subradiance is solved in TD basis, even in the absence of Lamb shift, there exists virtual couplings among TD states. These couplings arise fundamentally due to the interaction of discrete atomic energy states with a common environmental continuum. Such couplings were first studied by Ugo Fano in his 1961 seminal paper \cite{fano1961effects}. In the context of quantum statistical theories of spontaneous emission these types of couplings were first pointed out by Agarwal \cite{agarwal1974quantum}. Therefore, in view of \cite{scully2007correlated}, we'll refer such interactions as the Fano-Agarwal (FA) couplings.
\subsection{Decay of $\ket{+}_{\vec{k}_{0}}$ and $\ket{-}_{\vec{k}_{0}}$ states}
We start off with treating the full problem of N-atom sample prepared initially in either symmetric $\ket{+}_{\vec{k}_{0}}$ or first antisymmetric $\ket{-}_{\vec{k}_{0}}$ TD state. To analyze the influence of pure FA couplings we'll apply the RWA and neglect the presence of two excitations in the state of the system. Consequently, N-atom Hamiltonian takes the form:
\begin{equation}
\begin{split}
&\hat{V}=\hbar\sum_{j,k} g_{k}(\hat{a}_{k}\hat{\sigma}_{j}^{\dagger}e^{i(\omega-\nu_{k})t+i\vec{k}.\vec{r}_{j}}+\hat{a}^{\dagger}_{k}\hat{\sigma}_{j}e^{-i(\omega-\nu_{k})t-i\vec{k}.\vec{r}_{j}}).
\end{split}
\end{equation}
The system-environment state can be expressed as:
\begin{equation}
\begin{split}
&\ket{\Psi(t)}=\Bigg(\beta_{+}\ket{+}_{\vec{k}_{0}} + \beta_{-}\ket{-}_{\vec{k}_{0}} +...+\beta_{-}\ket{N}_{\vec{k}_{0}}\Bigg)\otimes \ket{0}\\
&+\sum_{\vec{k}}\gamma_{\vec{k}}(t)\ket{b_{1},b_{2},...,b_{N},1_{\vec{k}}},
\end{split}
\end{equation}
The general form of the antisymmetric TD states can be presented as \cite{svidzinsky2008dynamical}:
\begin{equation}
\begin{split}
&\ket{N}_{\vec{k}_{0}}=\frac{1}{\sqrt{N(N-1)}}\Bigg[\sum^{N-1}_{j=1}e^{i\vec{k}_{0}.\vec{r}_{j}}\ket{b_{1}, b_{2},..., a_{j},...b_{N}}\\
&-(N-1)e^{i\vec{k}_{0}.\vec{r}_{N}}\ket{b_{1}, b_{2}...a_{N}}\Bigg], \hspace{3mm}\forall N \geq 2
\end{split}
\end{equation}
while $\ket{N}_{\vec{k}_{0}}$ is equal to $\ket{-}_{\vec{k}_{0}},\ket{3}_{\vec{k}_{0}},...$ for $N=2,3,...$ respectively. Notice that the above choice of the structure of antisymmetric TD states is not unique but it can easily extendable to many atoms ensemble. Following only the evolution of $\beta_{\pm}(t)$ we obtain:
\begin{equation}
\begin{split}
&\dot{\beta}_{+}=\frac{-\gamma}{N}\Bigg[\sum_{i,j}e^{-i\vec{\mathcal{K}_{ji}}}\frac{sin\mathcal{K}_{ji}}{\mathcal{K}_{ji}}\beta_{+}+ \Bigg(\sum_{i,j}e^{-i.\vec{\mathcal{K}_{ji}}}\frac{sin\mathcal{K}_{ji}}{\mathcal{K}_{ji}}\\
&-\sum_{i^{'},j}e^{-i\vec{\mathcal{K}_{ji^{'}}}}\frac{sin\mathcal{K}_{ji^{'}}}{\mathcal{K}r_{ji^{'}}}\Bigg) \beta_{-}+...+N^{th}term\Bigg]\\
&\dot{\beta}_{-}=\frac{-\gamma}{N}\Bigg[\Bigg(\sum_{i,j}e^{-i\vec{\mathcal{K}_{ji}}}\frac{sin\mathcal{K}_{ji}}{\mathcal{K}_{ji}}\nonumber-\sum_{i^{'},j}e^{-i\vec{\mathcal{K}_{ji^{'}}}}\frac{sin\mathcal{K}_{ji^{'}}}{\mathcal{K}_{ji^{'}}}\Bigg)\beta_{+}\\
&+\Bigg(\sum_{i,j}e^{-i\vec{\mathcal{K}_{ji}}}\frac{sin\mathcal{K}_{ji}}{\mathcal{K}_{ji}}-\sum_{i',j}e^{-i\vec{\mathcal{K}_{ji'}}}\frac{sin\mathcal{K}_{ji'}}{\mathcal{K}_{ji'}} \\ 
& -\sum_{i,j'}e^{-i\vec{\mathcal{K}_{j'i}}}\frac{sin\mathcal{K}_{j'i}}{\mathcal{K}_{j'i}}+\sum_{i',j'}e^{-i\vec{\mathcal{K}_{j'i'}}}\frac{sin\mathcal{K}_{j'i'}}{\mathcal{K}_{j'i'}}\Bigg)\beta_{-}\\
&+...+N^{th}term\Bigg],
\end{split}
\end{equation}
while ${\mathcal{K}_{ji}}\equiv k_{0}r_{ji}$ and $\vec{\mathcal{K}_{ji}}=\vec{k}_{0}.\vec{r}_{ji}$. In order to further proceed with analytic results, we notice that the general structure of coupled differential equations for N atom case can be written in matrix form:
\begin{equation}
\begin{pmatrix}
\dot{\beta}_{+}(t)  \\
\dot{\beta}_{-}(t)  \\
  \vdots  \\
\dot{\beta}_{N}(t) 
\end{pmatrix}
=
\begin{pmatrix}
 \gamma_{11} & \gamma_{12} & \cdots & \gamma_{1N} \\
 \gamma_{21} & \gamma_{22} & \cdots & \gamma_{2N} \\
 \vdots  & \vdots  & \ddots & \vdots  \\
 \gamma_{N1} & \gamma_{N2} & \cdots & \gamma_{NN} 
\end{pmatrix}
\begin{pmatrix}
\beta_{+}(t)  \\
\beta_{-}(t)  \\
  \vdots  \\
\beta_{N}(t) 
\end{pmatrix}.
\end{equation}
 General solution of above equation can be represented as: 
\begin{equation}
\beta_{i}(t)=\sum^{N}_{i=1}c_{i}(t)V_{i}e^{\lambda_{i}t}, \hspace{2mm} i=+,-,3,...,N,
 \end{equation}
$V_{i}$ and $\lambda_{i}$ are the eigenvectors and eigenvalues of the effective decay rate matrix of Eq.~12. From this point onwards, analytic solutions are complicated to obtain without the imposition of further approximations (for example assuming extremely dense ensembles \cite{svidzinsky2010evolution}). We on the other hand want to keep the analysis more accurate here and proceed with numerical analysis of the problem where the position of all atoms will be treated discretely.
\begin{figure}[t]
\begin{center}
  \begin{tabular}{@{}cccc@{}}
     \includegraphics[width=2.6in, height=2in]{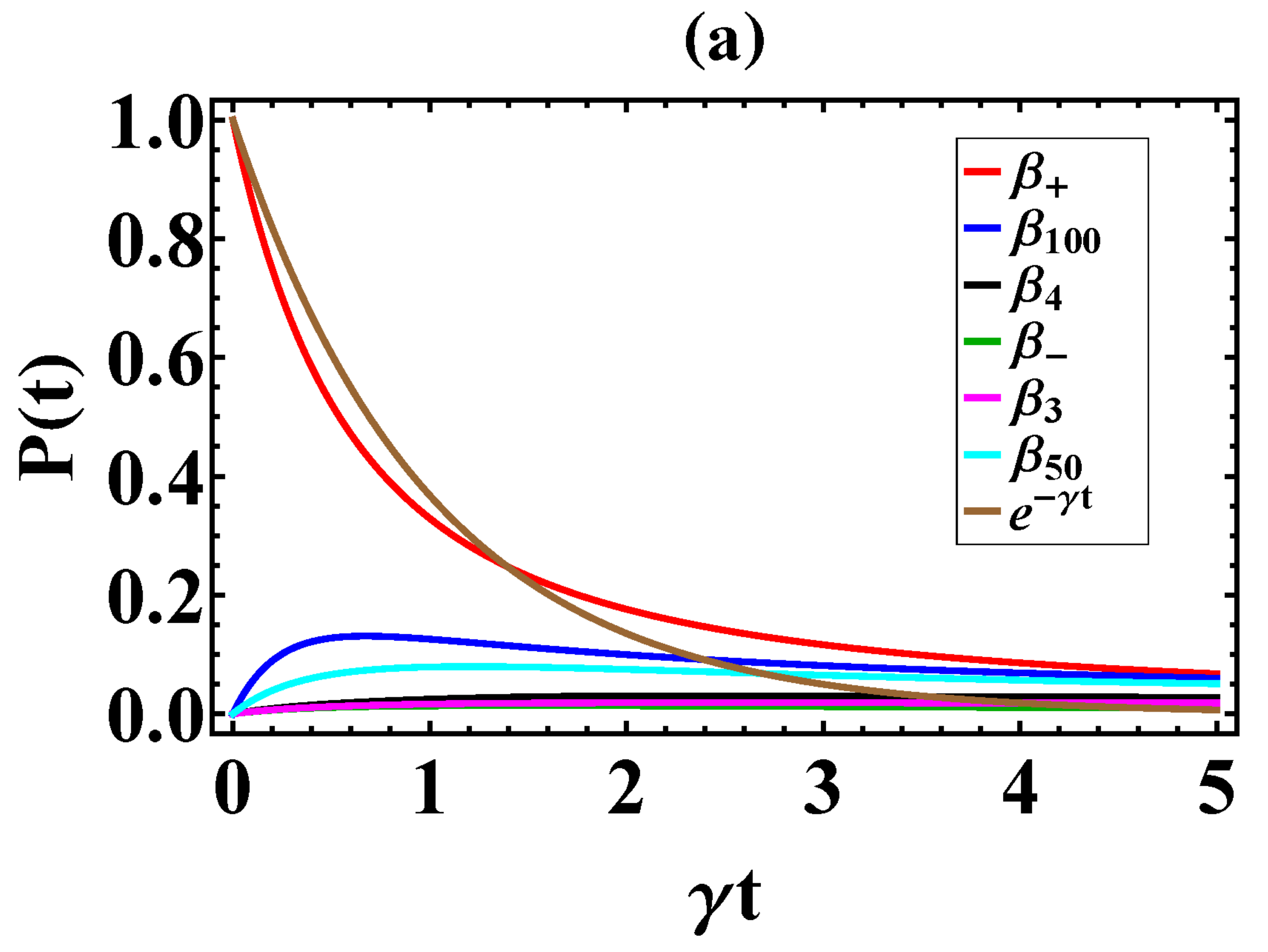} \\
   \includegraphics[width=2.6in, height=2in]{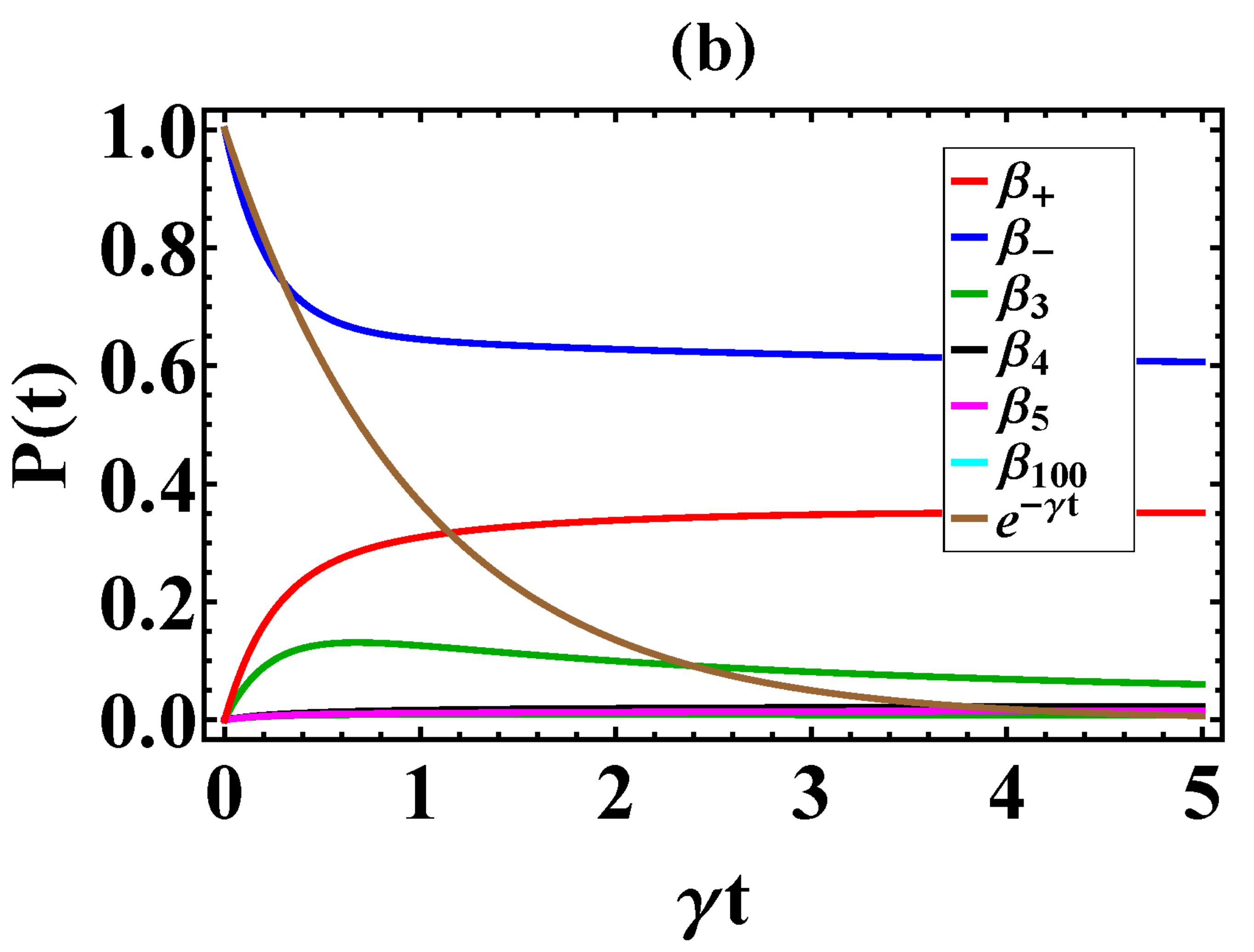}
  \end{tabular}
  \captionsetup{
  format=plain,
  margin=1em,
  justification=raggedright,
  singlelinecheck=false
}
   \caption{Time evolution of symmetric and antisymmetric TD states populations for an ensemble of 100 atoms in a line geometry. Part (a) ((b)) of plots represents the case when the system start off in a symmetric (antisymmetric) state. The lattice constant is $1k_{0}$ and radiation wavelength is $2\pi$ with $cos(\theta_{0})=1$.}\label{Fig1}
   \end{center}
\end{figure} 

To this end, we have performed the numerical simulation using the Runge-kutta method of order 4 with time step size $dt=0.01$ and all atoms periodically placed on a line lattice with lattice constant $1\lambda_{0}$ ($\lambda_{0}=2\pi/k_{0}$). In Fig.~1, we present the time evolution of populations for a hundred atom sample. When the system starts off in the symmetric state (Fig.~1(a)), we notice that the symmetric state shows slightly faster decay for smaller times ($\gamma t<2$) but still cannot be regarded as a superradiant state.\\
 The population in the antisymmetric states achieve small  maximum values, however the number of these curves also grow. Therefore, to a good approximation one can neglect the FA contribution coming just from $\beta_{-}$ in the evolution of $\beta_{+}$ (as analytically shown in the reference \cite{scully2015single}). However, the overall effect of all FA couplings is considerable and hence cannot be omitted. On the other hand, when the ensemble start off in the first antisymmetric state (Fig.~1(b)), the decay in the antisymmetric state population is slightly worsened. But in all cases the decay is subradiant.
\subsection{FA couplings between individual TD states}
The question of how an atomic ensemble initially prepared in the ordinary Dicke (OD) or symmetric TD state remains excited has already been investigated in past \cite{svidzinsky2009evolution}. In this subsection we direct our attention to two novel questions: (1) How the atomic ensemble decays in time if the initial preparation is in a TD subradiant state? (see next section for the analysis of this question in the presence of Lamb shift and NRW terms.)(2) During the decay process, how different FA couplings between certain TD state compare?\\
\begin{figure}[t]
\begin{center}
  \begin{tabular}{@{}cccc@{}}
     \includegraphics[width=1.6in, height=1.45in]{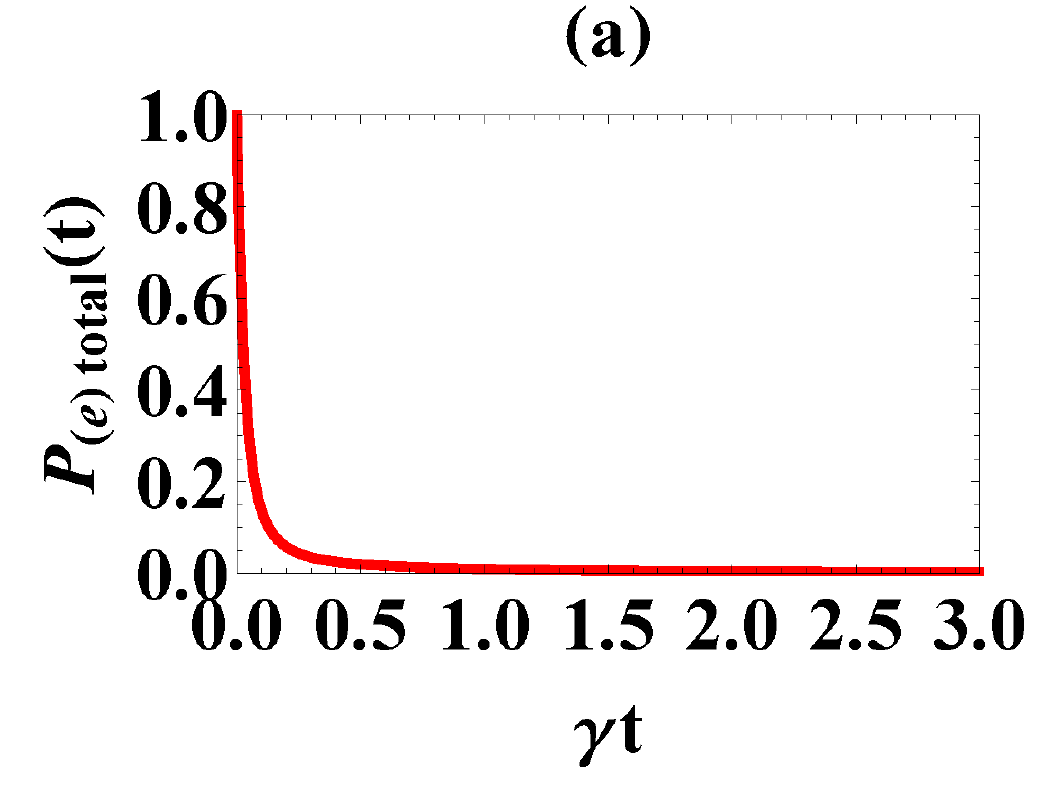}&
  \includegraphics[width=1.6in, height=1.5in]{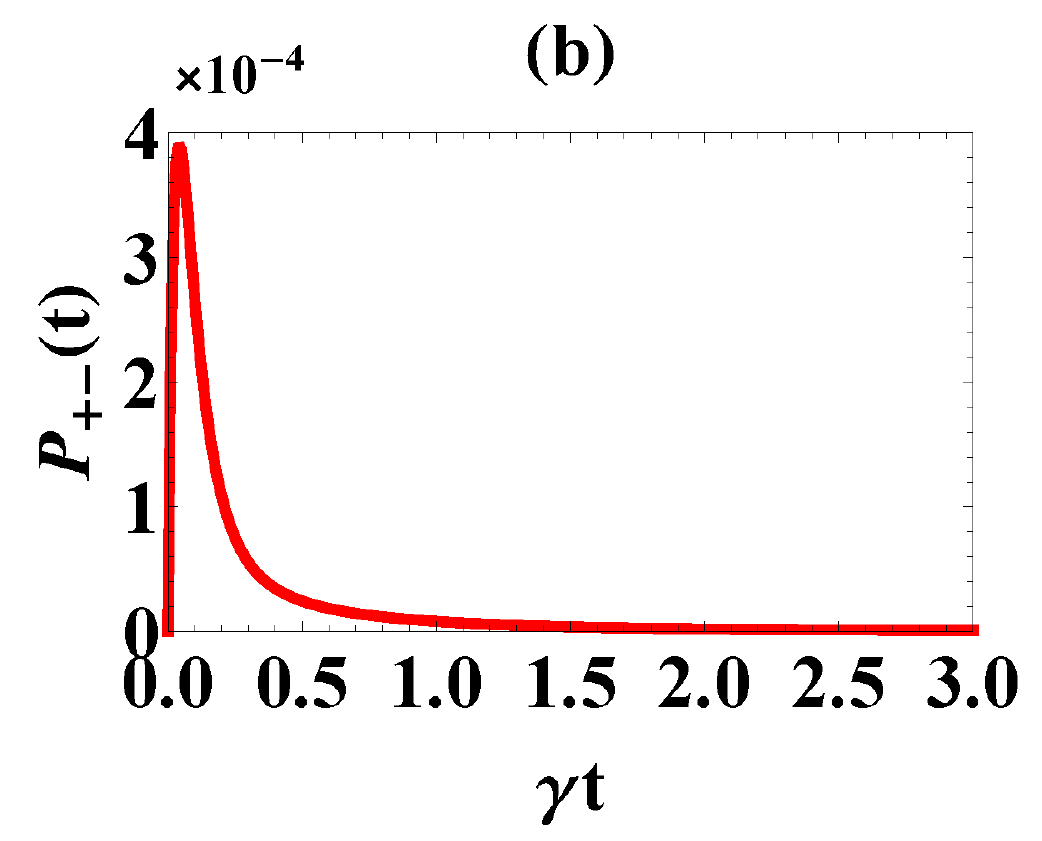}\\
  \includegraphics[width=1.6in, height=1.45in]{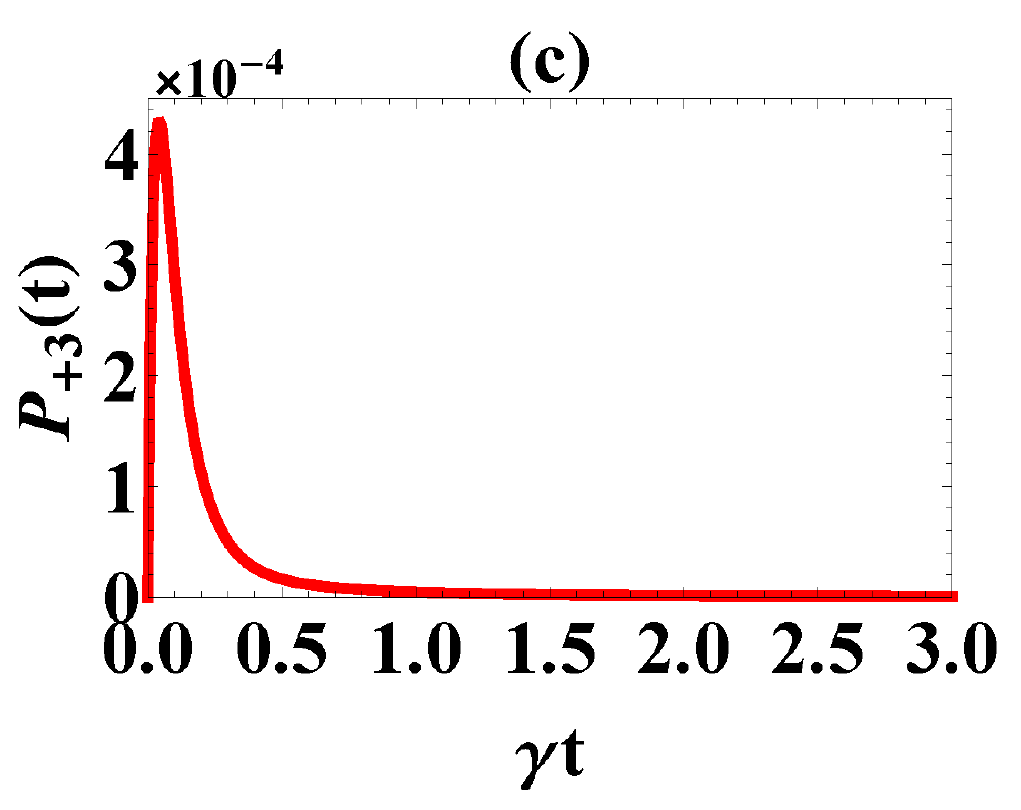}& 
  \includegraphics[width=1.6in, height=1.5in]{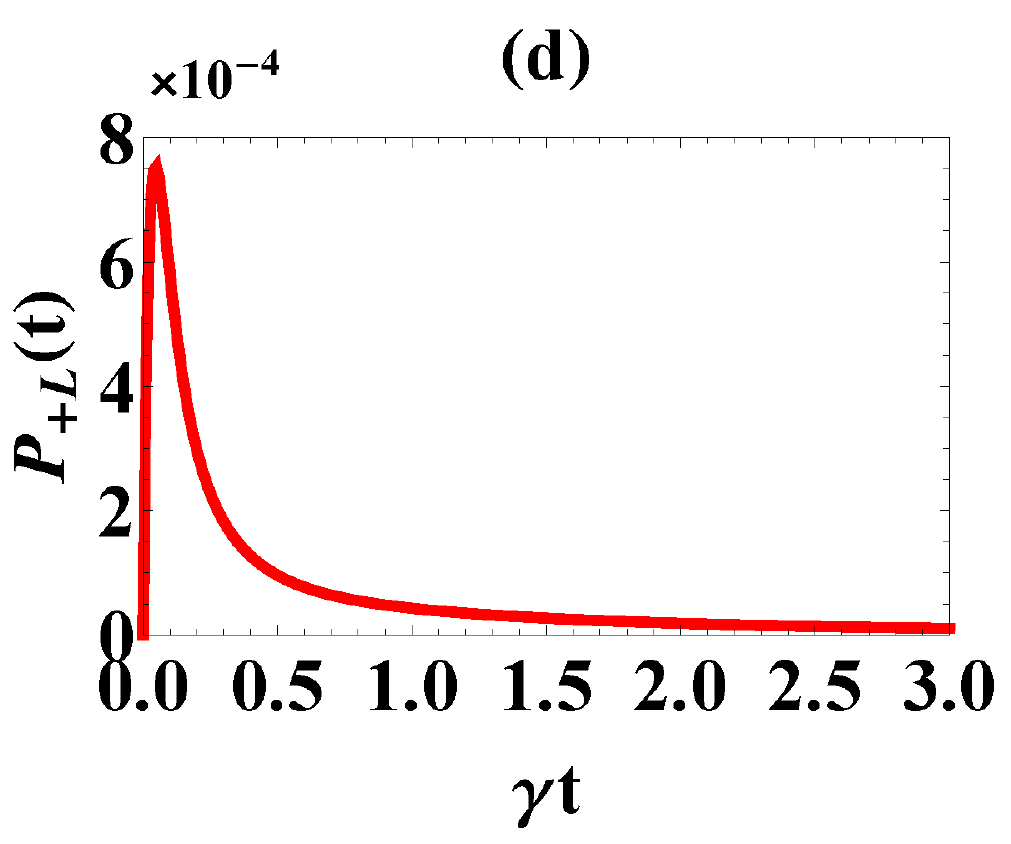}
  \end{tabular}
  \captionsetup{
  format=plain,
  margin=1em,
  justification=raggedright,
  singlelinecheck=false
}
   \caption{Time evolution of (a) full sample to be excited and FA coupling between: (b) $\ket{+}_{\vec{k}_{0}}$ and $\ket{-}_{\vec{k}_{0}}$ (c) $\ket{+}_{\vec{k}_{0}}$ and $\ket{3}_{\vec{k}_{0}}$ and (d) $\ket{+}_{\vec{k}_{0}}$ and $\ket{121}_{\vec{k}_{0}}$ states. System here starts off in the symmetric TD state. For this and next figure, we have conisered a spherical ensemble with radius $3k^{-1}_{0}$ and a uniform distribution of 121 atoms in the sphere. Inter-atomic separation is $1/k_{0}$ and the angle between the $\vec{k_{0}}$ and $\vec{r}_{ij}$ is decided by the coordinates of each $\vec{r}_{ij}$ while $\vec{k}_{0}=(1,0,0)$}\label{Fig2}
   \end{center}
\end{figure}
\begin{figure}[t]
\begin{center}
  \begin{tabular}{@{}cccc@{}}
     \includegraphics[width=1.6in, height=1.5in]{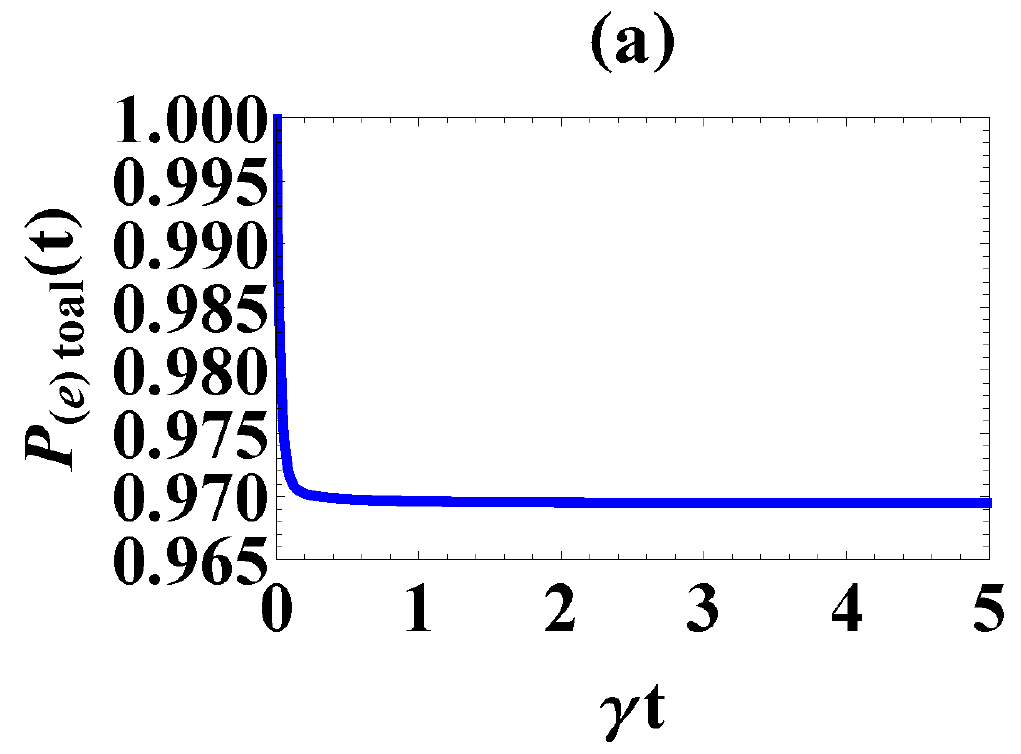}&
  \includegraphics[width=1.6in, height=1.4in]{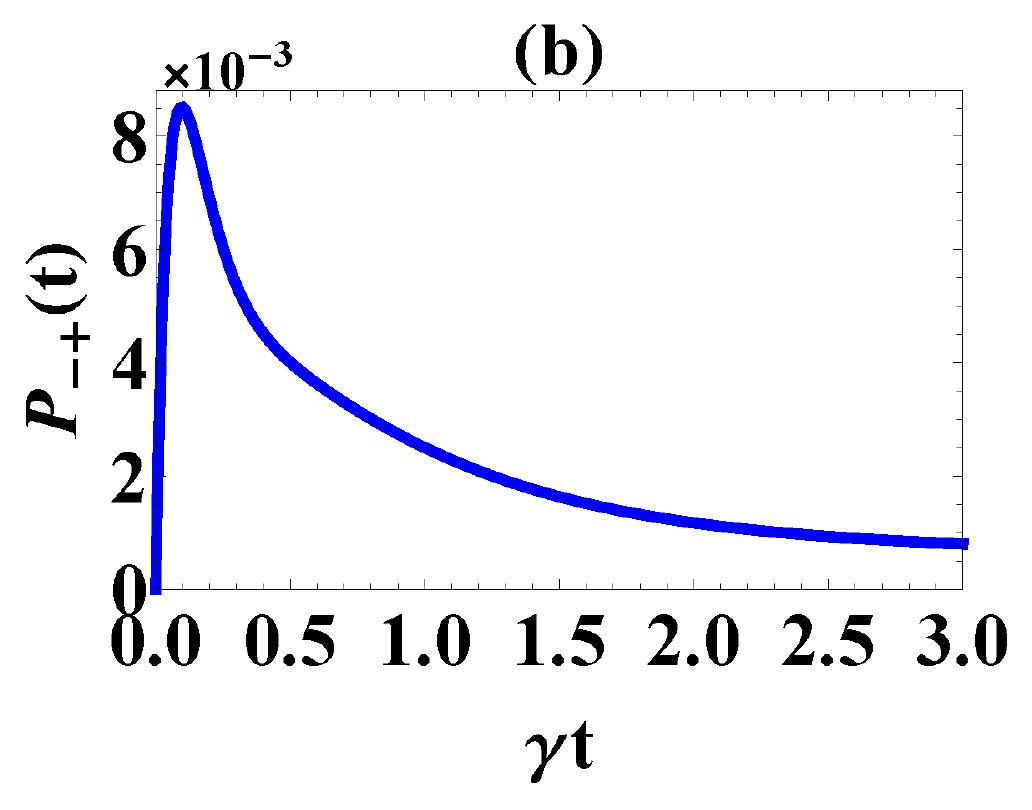}\\
 \includegraphics[width=1.6in, height=1.5in]{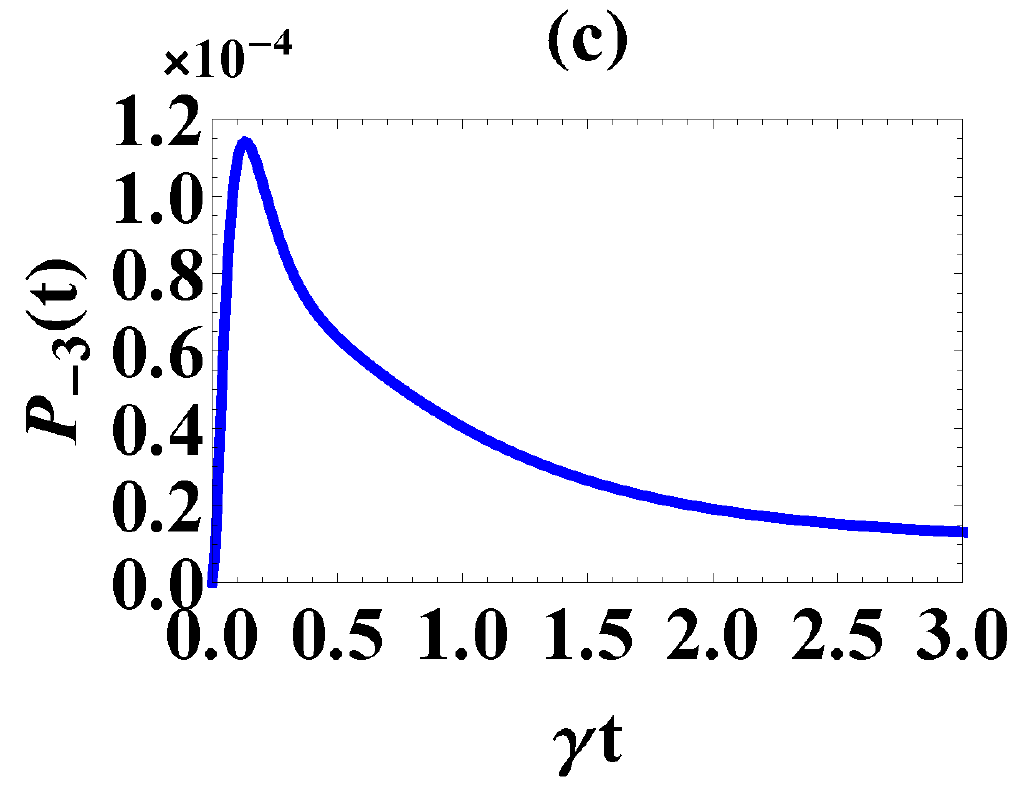}& 
  \includegraphics[width=1.6in, height=1.4in]{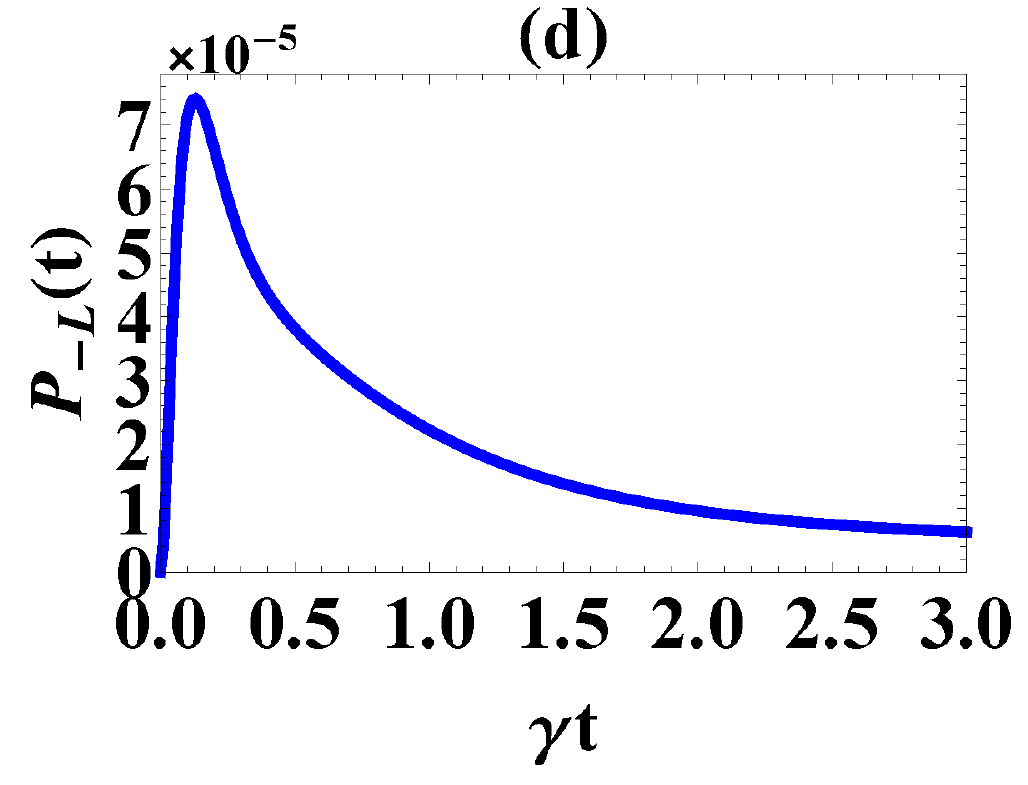}
  \end{tabular}
  \captionsetup{
  format=plain,
  margin=1em,
  justification=raggedright,
  singlelinecheck=false
}
   \caption{Time evolution of (a) full sample to be in an excited state and FA coupling between: (b) $\ket{-}_{\vec{k}_{0}}$ and $\ket{+}_{\vec{k}_{0}}$ (c) $\ket{-}_{\vec{k}_{0}}$ and $\ket{3}_{\vec{k}_{0}}$ and (d) $\ket{-}_{\vec{k}_{0}}$ and $\ket{121}_{\vec{k}_{0}}$ states, when system starts off in the $\ket{-}_{\vec{k}_{0}}$ TD state. }\label{Fig3}
   \end{center}
\end{figure}
 In Fig. 2(a) we plot the total probability of a spherical ensemble to remain excited if the initial state of the system is $\ket{+}_{\vec{k}_{0}}$. We notice an extremely fast (but not as fast as $N\gamma$) superradiant decay. In parts (b), (c) and (d) we have plotted the FA coupling between $\ket{+}_{\vec{k}_{0}}$ and $\ket{-}_{\vec{k}_{0}}$, $\ket{3}_{\vec{k}_{0}}$ and $\ket{121}_{\vec{k}_{0}}$ states, respectively. We notice FA couplings to be larger between $\ket{+}_{\vec{k}_{0}}$ and TD state with largest N. This can be understood noting that for $N>>1$, the desired coupling probability is proportional to: $P_{+\rightarrow N}\propto|\bra{+}_{\vec{k}_{0}}.(\sum^{N-1}_{j=1}\frac{e^{i\vec{k}_{0}.\vec{r}_{j}}}{N}\ket{j}-e^{i\vec{k}_{0}.\vec{r}_{N}}\ket{b_{1}b_{2}...a_{N}})|$. In this probability, we find that as N tends to achieve higher values, the FA coupling of $\ket{+}_{\vec{k}_{0}}$ state with larger N states start to enhance, as for these states the first term contribution surpasses second term contribution. \\

In Figure 3, we focus on the other scenario when the system starts off in $\ket{-}_{\vec{k}_{0}}$ TD state. Part (a) of the figure shows the net probability of system to remain excited. We notice a slight but fast decay of probability upto $t\leq 0.1\gamma^{-1}$ and after this time the probability shows an almost time independent behavior. In part (b), (c) and (d) we plot individual FA couplings of state $\ket{-}_{\vec{k}_{0}}$ state with $\ket{+}_{\vec{k}_{0}}$, $\ket{3}_{\vec{k}_{0}}$ and $\ket{121}_{\vec{k}_{0}}$ states respectively. We point out a contrary behavior of FA couplings in this case as compared to $\ket{+}_{\vec{k}_{0}}$ state situation (Fig.~2). Now $\ket{-}_{\vec{k}_{0}}$ maximally couples with $\ket{+}_{\vec{k}_{0}}$ state and as we go further away (i.e. higher N values in $\ket{N}_{\vec{k}_{0}}$) FA couplings become smaller.This trend can again be attributed to the desired coupling probabilities. There are two relevant probabilities now: (1) between symmetric and minus TD state $P_{-\rightarrow +}\propto|\bra{-}_{\vec{k}_{0}}.(\sum^{N-1}_{j=1}\frac{e^{i\vec{k}_{0}.\vec{r}_{j}}}{\sqrt{N}}\ket{j}|$ and (2) between minus and any other $N\geq 3$ subradiant state $\ket{N}_{\vec{k}_{0}}$ TD state which is proportional to:  $P_{-\rightarrow N}\propto|\bra{-}_{\vec{k}_{0}}.(\sum^{N-1}_{j=1}\frac{e^{i\vec{k}_{0}.\vec{r}_{j}}}{N}\ket{j}-e^{i\vec{k}_{0}.\vec{r}_{N}}\ket{b_{1}b_{2}...a_{N}})|$, for $N>1$. We note that if in the $\ket{-}_{\vec{k}_{0}}$, only atoms at position $\vec{r}_{1}$ and $\vec{r}_{2}$ participate in defining the antisymmetry then the second term in $\ket{N}_{\vec{k}_{0}}$ (in which the last (Nth) atom is excited) never participates in the desired probability. In this situation, as N becomes larger, the contribution from the first term in $P_{-\rightarrow N}$ becomes smaller as compared to the overlap of $\ket{-}_{\vec{k}_{0}}$ with $\ket{+}_{\vec{k}_{0}}$ state. Interestingly, we notice that even when two atoms that are defining the antisymmetry in $\ket{-}_{\vec{k}_{0}}$ state are placed at arbitrary positions in respective atomic bins (the situation actually plotted in Fig.~2 and 3), the same pattern of decay holds. 
\section{Inclusion of Lamb shift (Scalar theory) and non-rotating wave terms}
\begin{figure}
\includegraphics[width=3.4in, height=3.4in]{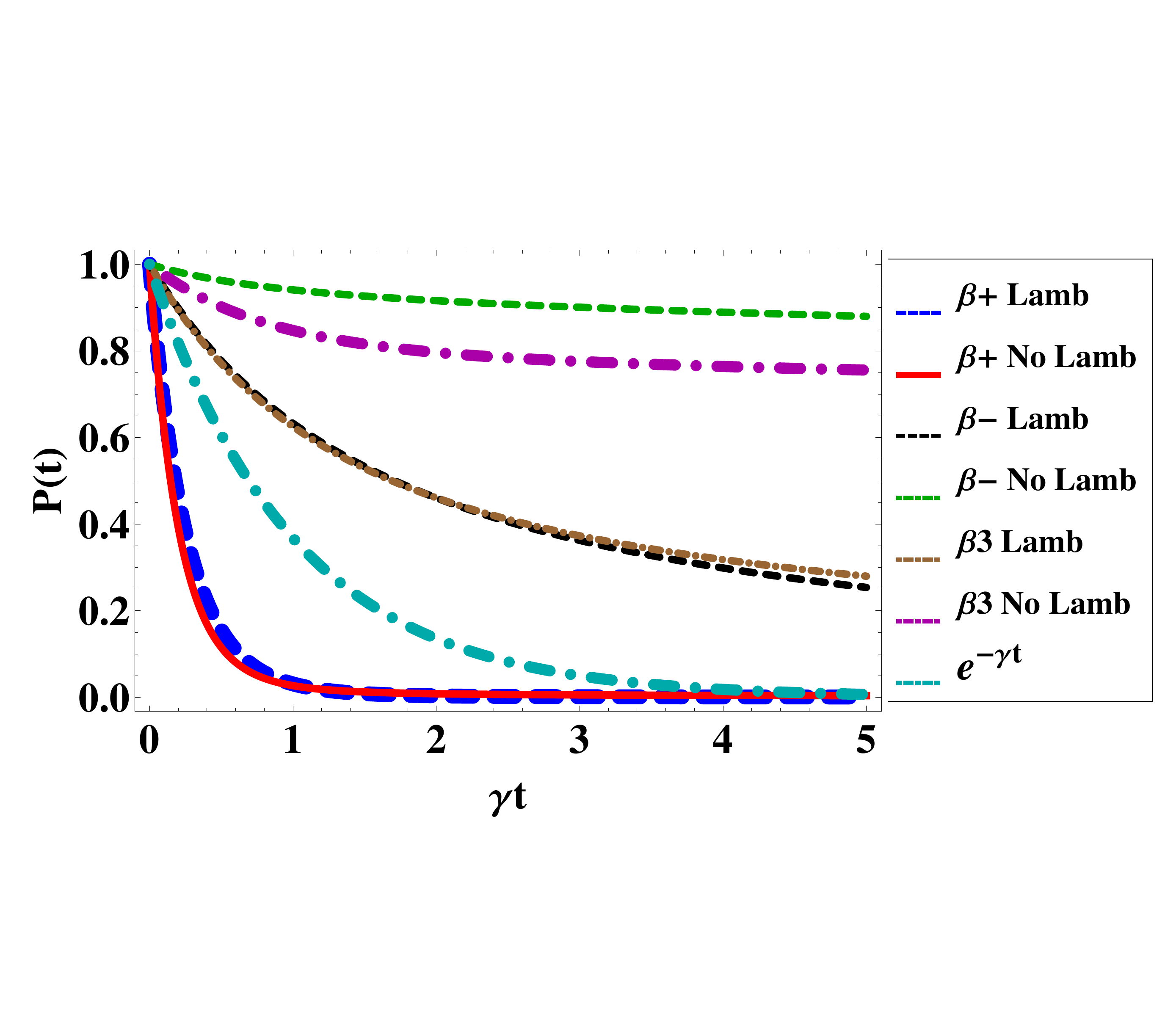}
\captionsetup{
 format=plain,
 margin=1em,
 justification=raggedright,
  singlelinecheck=false
}
  \vspace{-20mm}\caption{Lamb shift influencing the decay of an atomic ensemble initially prepared either in $\ket{+}_{\vec{k}_{0}}$ or $\ket{-}_{\vec{k}_{0}}$ or $\ket{3}_{\vec{k}_{0}}$ state. Here we consider an ensemble in spherical geometry with diameter $5\lambda_{0}$ ($\lambda_{0}$ is the wavelength of single photon in resonance with the atomic transition frequency). Ensemble consists of $1000$ periodically and uniformity distributed atoms with inter-atomic separation equal to $\lambda_{0}$. For $\ket{-}_{\vec{k}_{0}}$ and $\ket{3}_{\vec{k}_{0}}$ states we have divided the ensemble into two and three sections respectively, with each section having equal number of atoms. }\label{Fig4}
\end{figure}

We now examine the effect of Lamb shift and the non-rotating wave terms on the evolution of symmetric ($\ket{+}_{\vec{k}_{0}}$) and new kind of subradiant TD states ($\lbrace \ket{-}_{\vec{k}_{0}}, \ket{3}_{\vec{k}_{0}}, ..., \ket{N}_{\vec{k}_{0}} \rbrace$). We apply a scalar theory for simplicity here (for a comparison between a scalar and vector theory of electromagnetic modes decay from a spherical sample we refer \cite{friedberg2008electromagnetic}). We make use of the full Hamiltonian presented in Eq.~5 along with the global system-enviornment state presented in Eq.~6. The equations of motion for $\beta_{+}(t)$ and $\beta_{-}(t)$ amplitudes now takes the form:
\begin{equation}
\begin{split}
&\dot{\beta}_{+}=\frac{i\gamma}{N}\Bigg[\sum_{i,j}e^{-i\vec{\mathcal{K}_{ji}}}\frac{e^{i\mathcal{K}_{ji}}}{\mathcal{K}_{ji}}\beta_{+}+ \Bigg(\sum_{i,j}e^{-i\vec{\mathcal{K}_{ji}}}\frac{e^{i\mathcal{K}_{ji}}}{\mathcal{K}_{ji}}-\\
&\sum_{i^{'},j}e^{-i\vec{\mathcal{K}_{ji^{'}}}}\frac{e^{i\mathcal{K}_{ji^{'}}}}{\mathcal{K}_{ji^{'}}}\Bigg) \beta_{-}+...+N^{th}term\Bigg]\nonumber
\end{split}
\end{equation}
\begin{equation}
\begin{split}
&\dot{\beta}_{-}=\frac{i\gamma}{N}\Bigg[\Bigg(\sum_{i,j}e^{-i\vec{\mathcal{K}_{ji}}}\frac{e^{i\mathcal{K}_{ji}}}{\mathcal{K}_{ji}}-\sum_{i^{'},j}e^{-i\vec{\mathcal{K}_{ji^{'}}}}\frac{e^{\mathcal{K}_{ji^{'}}}}{\mathcal{K}_{ji^{'}}}\Bigg) \beta_{+}\\
&+\Bigg(\sum_{i,j}e^{-i\vec{\mathcal{K}_{ji}}}\frac{e^{i\mathcal{K}_{ji}}}{\mathcal{K}_{ji}}-\sum_{i',j}e^{-i\vec{\mathcal{K}_{ji'}}}\frac{e^{i\mathcal{K}_{ji^{'}}}}{\mathcal{K}_{ji'}}-\sum_{i,j'}e^{-i\vec{\mathcal{K}_{j'i}}}\\
&\times\frac{e^{i\mathcal{K}_{j^{'}i}}}{\mathcal{K}_{j'i}}+\sum_{i',j'}e^{-i\vec{\mathcal{K}_{j'i'}}}\frac{e^{i\mathcal{K}_{j^{'}i^{'}}}}{\mathcal{K}_{j'i'}}\Bigg)\beta_{-}+...+N^{th}term\Bigg].
\end{split}
\end{equation}
To arrive at these equations we have performed the integration by the method of contours, where we have set $k_{0}\rightarrow k_{0}+i\epsilon$ (while $\epsilon<<1$) and identified $\frac{1}{x\mp i\epsilon}={\bf P}[\frac{1}{x}]\pm i\pi\delta(x)$. Note the appearance of additional timing factors ($e^{-i\vec{k}_{0}.\vec{r}_{ij}}$ and similar primed exponential kernel) factors makes our results different then the one obtain through using the Fock state basis (see Eq.~2 in the introduction section).\\
We now present the results both with and without the presence of Lamb shifts. We notice that the effect of Lamb shift on the decay of symmetric (superradiant) state is small and the Lamb shift slightly slows down the decay. Our result here is consistent with reference \cite{svidzinsky2009evolution} where the effect of Lamb shift on collective decay of a spherical dense ensemble prepared initially either in $\ket{+}$ or $\ket{+}_{\vec{k}_{0}}$ state was investigated. \\
However, the effect of Lamb shift is marked on the decay of antisymmetric subradiant TD states (upto $\sim 40\%$ faster decay when Lamb shifts are included). This behavior, which to our knowledge has not been reported before, can be understood as a consequence of enhanced couplings between $\ket{-}_{\vec{k}_{0}},\ket{3}_{\vec{k}_{0}}$ and $\ket{+}_{\vec{k}_{0}}$ states in the presence of virtual processes. This feature points out that along with the FA couplings, now there are additional coupling channels available among the single and two excitation states through virtual interactions. A small enhanced coupling of fragile subradiant states with the superradiant state causes a marked effect on the decay of subradiant state as opposed to when the superradiant state is elevatedly coupled with the subradiant states. Hence, the effect of Lamb shift is more pronounced for the subradiant states.

\section{Conclusions}
The proposal of utilizing atomic ensembles prepared in single-photon subradiant states \cite{scully2015single, bienaime2012controlled} for quantum information storage purposes crucially relies on how these states decay. Consequently, in this paper we have investigated the effects of virtual processes on the time evolution of new kind of TD subradiant states introduced in reference \cite{scully2015single}. We concluded that, in the case of pure FA couplings the symmetric TD state decay tends to slow down without any superradiance while the subrradiant states remain no more frozen (unlike OD case). Additionally, the overall (summed up) effect of FA couplings remained substantial and hence cannot be neglected. The analysis of the individual FA couplings among different TD states revealed that if we start in $\ket{+}_{\vec{k}_{0}}$ state, then the FA couplings is highest with largest N TD state. On contrary, if the ensemble in prepared initially in $\ket{-}_{\vec{k}_{0}}$, then the coupling diminishes for TD states with larger N. Finally, the inclusion of scalar Lamb shift and NRW terms yields a small effect on the decay of $\ket{+}_{\vec{k}_{0}}$ state, however $\ket{-}_{\vec{k}_{0}}$ (and other TD subradiant states) shows up to 40\% swift decay. 
\acknowledgments
IMM would like to thank Marlan O. Scully, Robin Kaiser and Anatoly Svidzinsky for helpful discussions and suggestions on the manuscript. We gratefully acknowledge support of the National Science Foundation Grants PHY-1241032 (INSPIRE CREATIV) and EEC-0540832 (MIRTHE ERC). 

\bibliography{Article}
\end{document}